\documentclass[12pt]{article} 

\usepackage[utf8]{inputenc} 

\usepackage{geometry} 
\usepackage{setspace}
\usepackage{graphicx,url} 

\usepackage{booktabs} 
\usepackage{array} 
\usepackage{paralist} 
\usepackage{verbatim} 
\usepackage{amssymb}
\usepackage{amsmath}
\usepackage{textcomp} 

\usepackage{fancyhdr} 
\usepackage[nottoc,notlof,notlot]{tocbibind} 
\usepackage[titles]{tocloft} 

\title{\vspace{-8ex}\textbf{Observables of super-extremal black holes: \\[-1ex] challenging Cosmic Censorship to \\[-1ex] comprehend the Cosmological Constant}\vspace{1ex}}
\author{\vspace{-2ex}
\textbf{Dr.~Jenny Wagner} \\[1ex] 
thegravitygrinch@gmail.com \\[-1ex]
Bahamas Advanced Study Institute \& Conferences \\[-1ex]
4A Ocean Heights, Hill View Circle, Stella Maris, Long Island, The Bahamas \\[-1ex]
\url{https://thegravitygrinch.blogspot.com}}

\begin{document}
\maketitle
\thispagestyle{empty}
\begin{singlespace}
\vspace{-6ex}
\begin{center}
\textit{Essay written for the \\Gravity Research Foundation 2024 Awards for Essays on Gravitation.}
\end{center}
\end{singlespace}
\begin{figure}[h!]
\begin{center}
\includegraphics[width=0.26\textwidth]{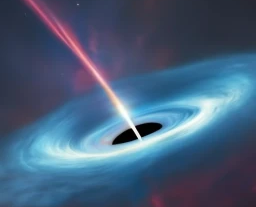}
\end{center}
\end{figure}
\vspace{-4ex}
\abstract{
\noindent 
Einstein's Field Equations have proven applicable across many scales, from black holes to cosmology. 
Even the mysterious Cosmological Constant found a physical interpretation in the so-called ``dark energy'' causing the accelerated cosmic expansion as inferred from multiple observables. 
Yet, we still lack a material source for this dark fluid.
Probing the local universe to find it yields complementary information to the one from the cosmic microwave background. 
Could dark energy be sourced by super-extremal charged black holes? 
Contrary to intuition, such objects could exist with only weak observational signatures. The latter are introduced here to outline how sky surveys can identify individual candidates which challenge Cosmic Censorship on the one hand but may explain the physical origin of the Cosmological Constant on the other.
} 

\clearpage
\pagenumbering{arabic} 
\subsection*{What is the Cosmological Constant?}

The Cosmological Constant, $\Lambda$, was first introduced into Einstein's field equations as a pragmatic remedy to prevent gravitational collapse of all matter and to maintain a static universe \cite{bib:Einstein1917}.
However, as discovered later, $\Lambda$ can also be used to describe an accelerated cosmic expansion \cite{bib:Perlmutter1999, bib:Riess1998}. 
Both interpretations for $\Lambda$ treat it as an effective field because both cosmological models are based on the Cosmological Principle, assuming a homogeneous and isotropic metric and matter sector on the largest cosmic scales and thereby representing cosmology as a Friedmann-Lemaître-Robertson-Walker (FLRW) model.
These interpretations of $\Lambda$ give the Cosmological Constant a precise meaning when an FLRW model is fitted to observational data \cite{bib:Ellis1984, bib:Ellis1987}.
Its meaning in other applications of Einstein's field equations, for instance describing black holes, remains to be explained.
As Einstein's Field Equations are applicable on many scales in the same way, one may also hope for a unifying interpretation of $\Lambda$ on all scales. 

A possibility to derive $\Lambda$ from fundamental principles was pursued in quantum field theory. 
Attempting to explain $\Lambda$ by quantum fluctuations of the zero-point energy of spacetime, the estimated theoretical value exceeded observational limits by some 120 orders of their magnitude \cite{bib:Abbott1988, bib:Weinberg1989}. 
\cite{bib:Nobbenhuis2006} describe this discrepancy as the smallness problem of $\Lambda$. 
They provide various, yet still insufficient explanations.

Since $\Lambda$ was introduced on the classical level, an understanding should also be gained classically. 
We pursue this approach by first motivating a possible interpretation as a gauge constant which fixes the reference amplitude of inert background energy-momentum density that does not cause space-time curvature. 
Subsequently, we investigate if at least part of the 69\% of the overall cosmic background energy-momentum content attributed to $\Lambda$ could be contained in super-extremal black holes. 
Such objects challenge the Cosmic Censorship conjecture \cite{bib:Penrose1998}, and are suspected not to exist \cite{bib:Wald1974}. 
A closer inspection of their observable signatures reveals, however, that they only leave mild imprints instead of disruptive events and therefore might be lurking in the local universe. 
Given the vague knowledge of cosmic times before the microwave background decoupling, possible generation channels could be created during electro-weak symmetry breaking.


\subsection*{Gauging Einstein's Field Equations}

Einstein's field equations in their canonical form without $\Lambda$ read 
\begin{equation}
{\rm{G}}_{\mu \nu} \equiv {\rm{R}}_{\mu \nu} - \frac{R}{2} {\rm{g}}_{\mu \nu} = \kappa {\rm{T}}_{\mu \nu}  \;,
\label{eq:EFE1}
\end{equation}
with the usual Einstein tensor ${\rm{G}}_{\mu \nu}$ consisting of the Ricci tensor ${\rm{R}}_{\mu \nu}$, $\mu, \nu = 0, ..., 3$ and the Ricci-scalar $R = {\rm{R}}^\mu_\mu$, depending on the metric ${\rm{g}}_{\mu \nu}$, the constant $\kappa = 8\pi G$, and the energy-momentum tensor ${\rm{T}}_{\mu \nu}$.
The left-hand side represents the dynamics of the metric sector and the right-hand side the one of matter. 
Hence, \eqref{eq:EFE1} implicitly assumes a separate dynamics of the metrics and the matter and establishes a relationship between the two by equating both sides. 
This separation becomes more obvious when noting that the divergences of both sides vanish individually.
For the left-hand side, \cite{bib:Lovelock1971} showed that the Einstein tensor ${\rm{G}}_{\mu \nu}$ and the metric itself ${\rm{g}}_{\mu \nu}$ are the only divergence-free tensors compatible with the symmetry requirements detailed in \cite{bib:Lovelock1971}. 
Reformulating the matter-metric-relation in terms of an action principle, \cite{bib:Schuller2018} derived the causally consistent action of the metric given the action of the matter.
Based on a few constraints on the spacetime metric, the energy-momentum content thus uniquely constrains the canonically compatible action of the metric. 
The approach recovers the Einstein-Hilbert action and obtains $\Lambda$ and the gravitational constant $G$ as the two remaining degrees of freedom not constrained by the matter content and its dynamics.

If $\Lambda$ is inserted into \eqref{eq:EFE1} the latter can also be rewritten as
\begin{equation}
{\rm R}_{\mu \nu} = \kappa {\rm{T}}_{\mu \nu} + \left( \Lambda - \frac{\kappa T^\lambda_\lambda}{2} \right) {\rm{g}}_{\mu \nu} \;.
\label{eq:Rmn}
\end{equation}
The physical meaning of the Ricci tensor on the left-hand side is to represent the leading-order deviation of a volume element in a curved spacetime, ${\rm d}V$, from a volume element in an Euclidean spacetime, ${\rm d}V_\mathrm{E}$, because 
\begin{equation}
{\rm d} V = \sqrt{{\rm det}({\rm{g}}_{\mu \nu})} \, {\rm d}V_\mathrm{E} \approx \left(1 - \frac16 {\rm{R}}_{\mu \nu} x^\mu x^\nu \right)  {\rm d}V_\mathrm{E}  \;,
\label{eq:dV}
\end{equation}
using Einstein's summing rule to contract the Ricci tensor with the coordinate vectors $x^\mu$.
Thus, the volume change due to curvature in \eqref{eq:dV} is determined by the right-hand side of \eqref{eq:Rmn}.
Adding $\Lambda$ to the reference value for ${\rm T}_{\mu \nu}$, $T^\lambda_\lambda$, rescales the latter and thereby fixes the curvature caused by the matter sector at the spacetime point under consideration. 

Altogether, $\Lambda$ represents the zero-point reference value which separates a homogeneous and isotropic background matter density from the perturbation-level matter that causes curvature. 
The gravitational constant $G$ determines how strong the curvature effects for a given amount of energy-momentum density are.
Thus, the unifying interpretation of $\Lambda$ at all classical scales amounts to being the global degree of freedom that sets the embedding background energy-momentum density.  
Assuming a specific metric and constraints on the overall energy-momentum content, $\Lambda$ can be constrained by inserting observations into \eqref{eq:Rmn}. 

To test this interpretation in the late-time, local universe, recent ideas to constrain $\Lambda$ from binary systems like Milky Way and Andromeda, \cite{bib:Benisty2023, bib:Benisty2024}, can be employed:
As long as calibrations by $N$-body simulations perfectly account for the surrounding masses, $\Lambda$ will remain compatible with being zero. 
Such a result means that $\Lambda$ represents the embedding background vacuum for this specific case, if ${\rm T}^\mu_\mu$ accounts for the visible and dark matter background density as assumed to be contained in the system.
With future measurements, the precision required to investigate whether $\Lambda$ is actually zero will be reached. 

On cosmological scales, a small, non-vanishing $\Lambda$ is significantly preferred by a multitude of data. 
Hence, we are missing mass on these scales. 
As a globally constant $\Lambda$-density on large scales is obtained by coarse-graining mass clumps on smaller scales, near-field cosmology as mentioned above can contribute to explore possible material sources. 
The latter must explain an isotropic accelerated expansion of the universe, but must have eluded our sky surveys so far.  
Attributing the missing mass to black holes has recently been discussed \cite{bib:Farrah2023, bib:Frampton2023}. 
Among the controversies, it is still unclear how such objects and suitable distributions can be created within our current concordance cosmology.
While this issue is explored by early-universe physicists, we now investigate if the existence of such objects is compatible with current observations and which future data can detect such objects.

\subsection*{Stupendously charged and massive primordial black holes}

We explore observable signatures of non-rotating super-extremal Reissner-Nordstr\"{o}m (RN) black holes. 
They may be of primordial origin, formed around the epoch of electro-weak symmetry breaking, when charges were generated at massive overdensities and immediately locked up into a naked singularity. 
On the smallest scales, electrons have a charge-to-mass-ratio to qualify as such objects, too, as suggested in \cite{bib:Einstein1938}.  
Primordial naked singularities avoid the issues raised by \cite{bib:Wald1974} that sub-extremal black holes cannot accrete charged, massive particles to turn into naked singularities at later times.
Moreover, they may persist over long times as it remains unclear if they can evaporate via Hawking radiation like their sub-extremal counterparts.
Thus, the Cosmic Censorship conjecture is worth being re-investigated.
The best way to do so is deriving observable signatures to detect objects violating it. 

In \cite{bib:Wagner2024}, we performed a proof-of-concept analysis: three example cases of super-extremal RN black holes with the properties summarised in Table~\ref{tab:cases} were investigated. 
Each case obeys $r_\mathrm{Q} > 1.5/\sqrt{8} \, r_\mathrm{S}$, i.e.~these objects have neither an event horizon nor a photon sphere. 
We also assumed that these objects repelled each other due to their charge at a distance of four megaparsecs. 
Induced electro-magnetic observables can then be constrained in the weak-gravitational field approximation using Li\'enard-Wiechert potentials for moving point charges.
Only the strong gravitational lensing signatures require strong-field calculations. 
Due to the Mpc-scale distances assumed, the inferred properties are independent of the cosmological background.
This is an advantage because we want to partly create the latter with these super-extremal black holes at a later stage. 
\vfill
\begin{table}[b!]
\begin{center}
\begin{tabular}{cccccccccc}
\textrm{case} & mass & mass $M$ & charge $Q$ & $r_\mathrm{S}$ & $r_\mathrm{Q}$  & $r_\mathrm{min}$ & $a$ & $\delta t$ \\ 
 & scale & $\left[ M_\odot \right]$ & $\left[ {\rm{C}} \right]$ & $\left[ \mbox{pc} \right]$ & $\left[ \mbox{pc} \right]$ & $\left[ \mbox{kpc} \right]$ & $\left[ {\rm{m}} {\rm{s}}^{-2} \right]$ & $\left[ a \right]$  \\
  \noalign{\smallskip}
 \hline 
 \noalign{\smallskip}
1 & galaxy & $10^{12}$ & $-4 \times 10^{32}$ & 0.10 & 0.1 & 0.01 & $10^{-14}$ & $10^7$ \\
2 & gal.~group & $10^{13}$ & $-4 \times 10^{34}$ & 1.0 & 10 & 0.1 & $10^{-11}$ & $10^5$ \\
3 & gal.~cluster & $10^{14}$ & $-4 \times 10^{36}$ & 10 & 1000 & 10 & $10^{-8}$ & $10^4$\\
\noalign{\smallskip}
\end{tabular}
\caption{\label{tab:cases} Specifications for example RN naked singularities with approximated values for the Schwarzschild radius $r_\mathrm{S}=2GM/c^2$ and $r_\mathrm{Q}=\sqrt{G/(4\pi\epsilon_0)}Q/c^2$, minimum distance $r_\mathrm{min}$ for the weak-field approximation to hold, acceleration $a$ due to mutual repulsion with a counterpart of the same specifications and time $\delta t$ needed to move by 2$r_\mathrm{Q}$.}
\end{center}
\end{table}

\begin{figure}[ht!]
\centering
\raisebox{7ex}{\includegraphics[width=0.43\textwidth]{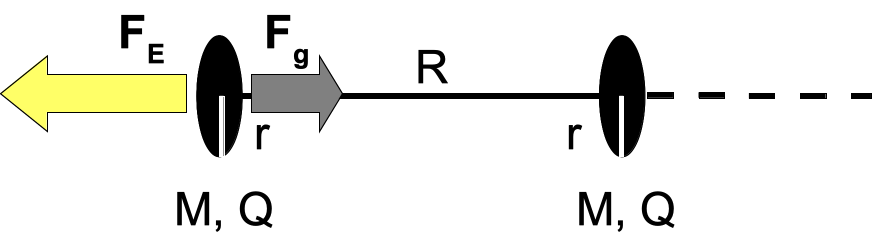}} \hspace{6ex}
\includegraphics[width=0.43\textwidth]{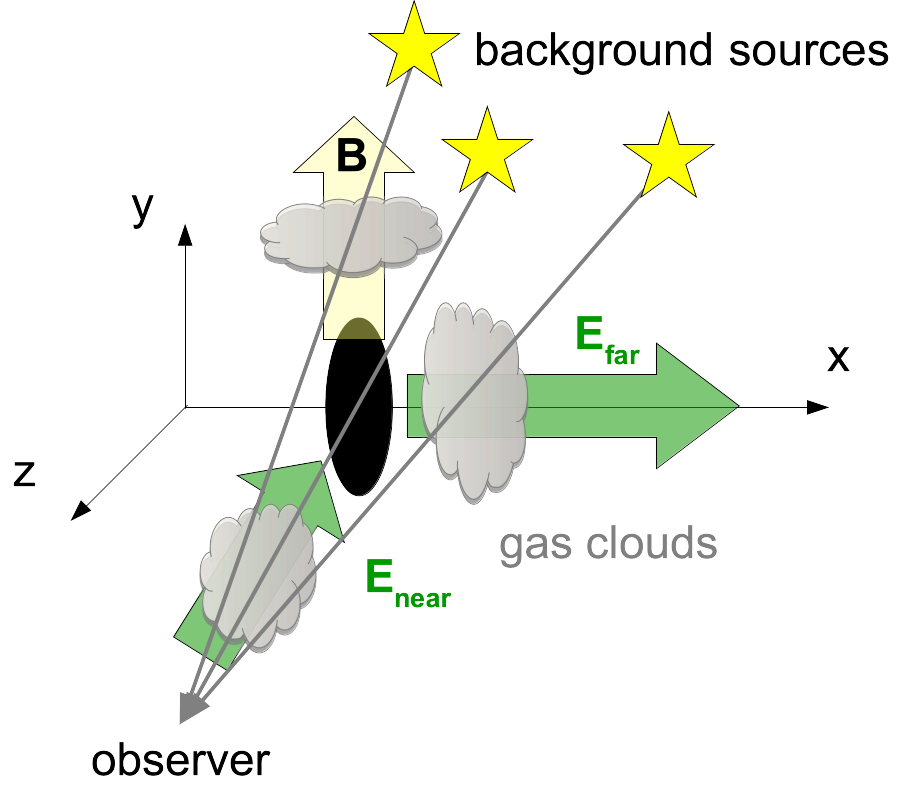}
\caption{Left: Mutual net repulsion of two super-extremal RN black holes as the Coulomb force $\boldsymbol{F}_\mathrm{E}$ exceeds gravitation $\boldsymbol{F}_\mathrm{g}$. Right: Induced Li\'enard-Wiechert fields for close-by gas clouds probed by emission and absorption spectroscopy, see text for details.}
\label{fig:scenario}
\end{figure}

Accelerations due to net mutual repulsion of two objects with the same characteristics and the time it takes to move by $2r_\mathrm{Q}$ out of rest are also listed in Table~\ref{tab:cases}.
They show that the induced electro-magnetic fields are static for the time of observations, but time-dependent on cosmic time scales. 
Fig.~\ref{fig:scenario} sketches the repulsion (left) and an induced electro-magnetic field configuration for one repelled super-extremal black hole moving along the $x$-axis, assuming the observer is located along the $z$-axis (right). 
While the electric field has near- and far-field parts in different directions, the magnetic field is aligned with the $y$-axis.
We align the gas clouds with the induced fields, such that the field strengths are maximal and then calculate the impact on spectrocsocpic signatures for varying distances. 
\vfill
\begin{figure}[hb!]
\centering
\includegraphics[width=0.45\textwidth]{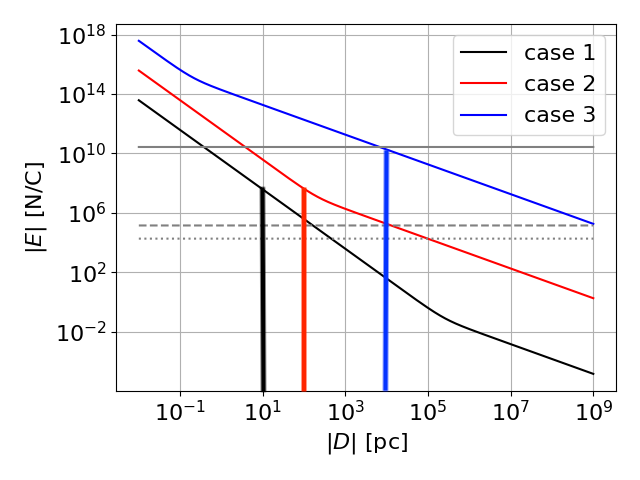} \hspace{6ex}
\includegraphics[width=0.45\textwidth]{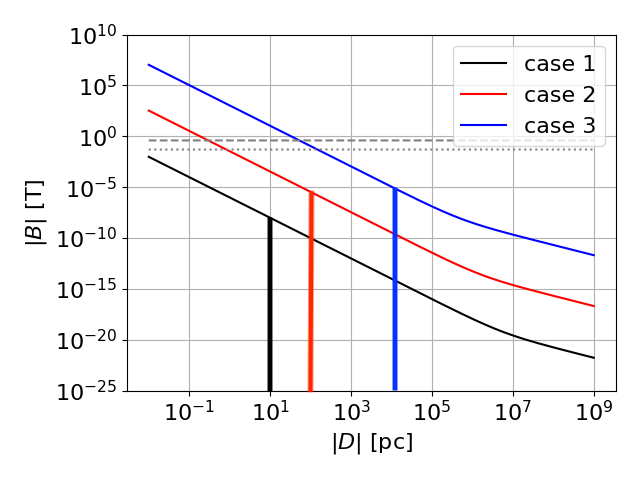}
\caption{Left: Maximum total induced electrical field strength versus distance of the gas cloud, vertical coloured bars denote $r_\mathrm{min}$ below which the weak-field approximation is invalid, horizontal thick grey lines denote b) and c) from the text. Right: Analogous plot for the maximum total induced magnetic field.}
\label{fig:fields}
\end{figure}

To investigate their effects on neutral or charged gas clouds, we compare the induced fields at distances larger than $r_\mathrm{min}$ to a) the electric field inducing proton decay $\approx 10^{13}$N/C \cite{bib:Wistisen2021}, b) the field required to dissociate neutral hydrogen $\approx 10^{10}$N/C, c) the minimum linear Stark effects to alter the fine and hyperfine structure of hydrogen $\approx 10^{5}$N/C and $\approx 10^{4}$N/C, respectively. 
Plotting the maximum electric fields to be expected for an optimal alignment of the gas cloud, we read off Fig.~\ref{fig:fields} (left) that none of our super-extremal RN black holes can induce proton decay, nor dissociate neutral hydrogen. 
The Stark effects, however, may perturb the fine and hyperfine structure signatures. 

Correspondingly, we determine the magnetic field effects on the expected spectra of gas clouds compared to the minimum amplitude of an external Zeeman effect to split the fine and hyperfine structure of neutral hydrogen gas $0.4$T and $0.05$T, respectively. 
As shown in Fig.~\ref{fig:fields} (right), even optimal alignments causing maximum magnetic field amplitudes are too weak to perturb the hydrogen spectra. 

While neutral hydrogen clouds only give weak hints at super-extremal RN black holes, plasma clouds may yield more prominent signals.
Linearly polarised radiation traversing a cloud of thermal electrons in an external magnetic field is subject to Faraday rotation and the observed rotation measure is modelled as
\begin{equation}
{\rm{RM}} = \left( \dfrac{8.37 \times 10^{-3}}{\mbox{m}^2} \right) \int \limits_{\boldsymbol{D}_\mathrm{i}}^{\boldsymbol{D}_\mathrm{f}} \left( \dfrac{n_e(\boldsymbol{s})}{1~\mbox{cm}^{-3}} \right)\left( \dfrac{\boldsymbol{B}}{1~\mbox{T}} \right)\cdot \left(\dfrac{\mathrm{d} \boldsymbol{s}}{1~\mbox{pc}} \right) \;,
\end{equation}
where the traversed length through the cloud has $\boldsymbol{D}_\mathrm{i}$ and $\boldsymbol{D}_\mathrm{f}$
 as end points, $n_\mathrm{e}(\boldsymbol{s})$ is the electron density at position $\boldsymbol{s}$, and $\boldsymbol{B}$ denotes the line-of-sight magnetic field induced by the naked singularity. 
 To show that this effect can be significant, we compare the rotation measures caused by our objects with those caused by other magnetic fields. 
 Table~\ref{tab:rm} summarises magnetic fields of various other sources and shows how far away the plasma cloud needs to be located from one of our example objects to have the same field strength, again assuming optimal alignment between magnetic field, plasma cloud, and observer for maximum field strength. 
 
\begin{table*}
\begin{center}
\begin{tabular}{ccccc} 
\textrm{scale} & $|\boldsymbol{B}|$ & $\left| \boldsymbol{D} \right|$ \textrm{case 1} & $\left| \boldsymbol{D} \right|$ \textrm{case 2} & $\left| \boldsymbol{D} \right|$ \textrm{case 3} \\ 
 & $\left[ \mbox{T} \right]$ & $\left[ \mbox{pc} \right]$ & $\left[ \mbox{kpc} \right]$ & $\left[ \mbox{Mpc} \right]$ \\ 
  \noalign{\smallskip}
 \hline 
 \noalign{\smallskip}
cosmic & $10^{-18}$ & $10^{6}$ & $> 10^{6}$ & $> 10^{3}$\\ 
\noalign{\smallskip}
galaxy & $10^{-9}$ - $10^{-8}$ & $10$ - $30$& $2$ - $6$ & $0.4$ - $2.5$ \\ 
\noalign{\smallskip}
star-burst region & $10^{-8}$ & $10$ & $2$ & $0.4$ \\ 
\noalign{\smallskip}
dense gas cloud & $10^{-8}$ - $10^{-7}$ & $< 10$ & $0.6$ - $2$ & $0.1$ - $0.4$ \\
\noalign{\smallskip}
\end{tabular}
\caption{\label{tab:rm}Comparison of magnetic fields on different scales in the universe with those maximally induced at distances $|\boldsymbol{D}|$ away from the three examples of Table~\ref{tab:cases}.}
\end{center}
\end{table*}

\subsection*{Strong-field strong lensing by naked singularities}

\begin{figure}[ht!]
\centering
\includegraphics[width=0.45\textwidth]{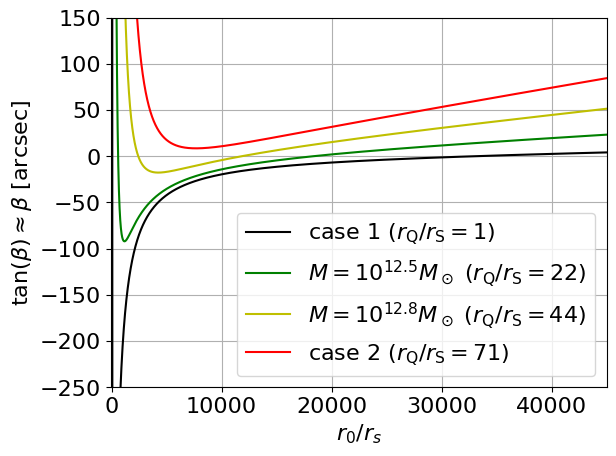} \hfill
\includegraphics[width=0.45\textwidth]{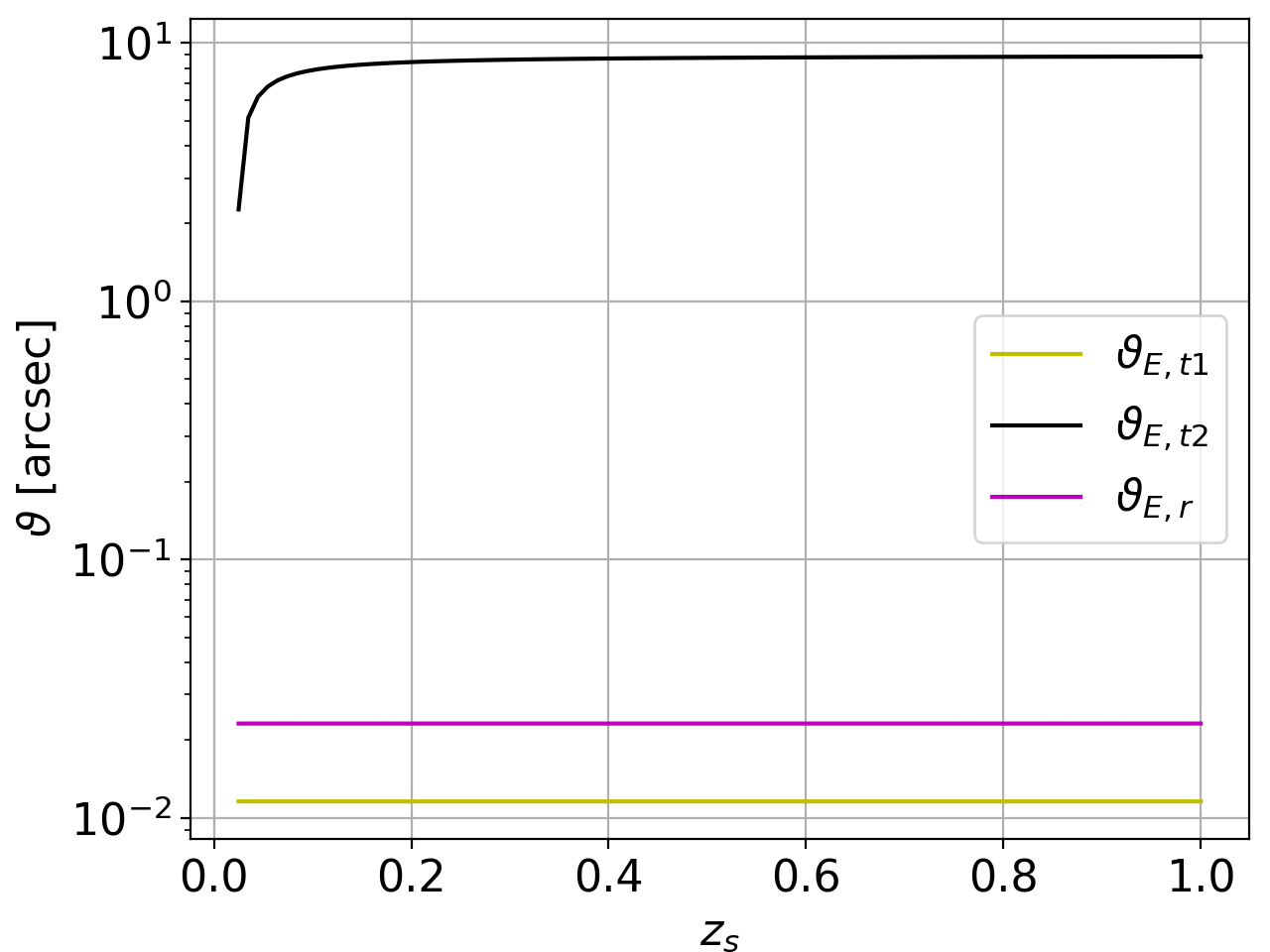}
\caption{Left: Source position $\beta$ w.r.t.~the lens centre vs.~impact parameters of deflected light for RN lenses with different $r_\mathrm{Q}/r_\mathrm{S}$-ratios. Right: Size of Einstein rings $\vartheta_\mathrm{E,t}$ and giant arc $\vartheta_\mathrm{r}$ occurring for case~1 in Table~\ref{tab:cases} as a lens at a distance of 100Mpc.}
\label{fig:rings}
\end{figure}

The most promising signature of super-extremal RN black holes is obtained when background light is deflected by the gravitational lensing effect. 
Due to the strongly singular nature of these objects -- they lack event horizons and photon spheres -- light at any impact parameter is deflected and nothing is absorbed.
In \cite{bib:Wagner2024}, we also determined the observables of the gravitational lensing effect in the full general relativistic framework. 
The latter is necessary because treating the singularity as a point-mass less in the weak-field approximation yields outrageously large Einstein rings. 
In the strong-field limit, however, both charge and mass contribute to the lensing effect, such that the outrageously large Einstein rings disappear above a certain $r_\mathrm{Q}/r_\mathrm{S}$-ratio.
Fig.~\ref{fig:rings} (left) shows this transition putting a naked singularity at 100Mpc distance as lens and a light source at redshift $z_\mathrm{s}=0.04$: Einstein rings occur for impact parameters $r_0/r_\mathrm{S}$ where the source position is exactly behind the lens centre, $\beta=0$.
Giant arcs occur at the extremum points of $\tan(\beta)\approx \beta$. 
Consequently, only radial arcs can be observed for cases~2 and 3, which is compatible with current observational data, lacking evidence for large Einstein rings.
Besides the arcsecond-scale Einstein ring for case~1, a peculiarity of this case is an additional Einstein ring at subarcsecond-scale that does not occur in the weak-field limit. 
Hence, we either obtain two Einstein rings and a giant arc, as depicted for case~1 in Fig.~\ref{fig:rings} (right), or just giant radial arcs. 

\begin{figure}[ht!]
\centering
\includegraphics[width=0.45\textwidth]{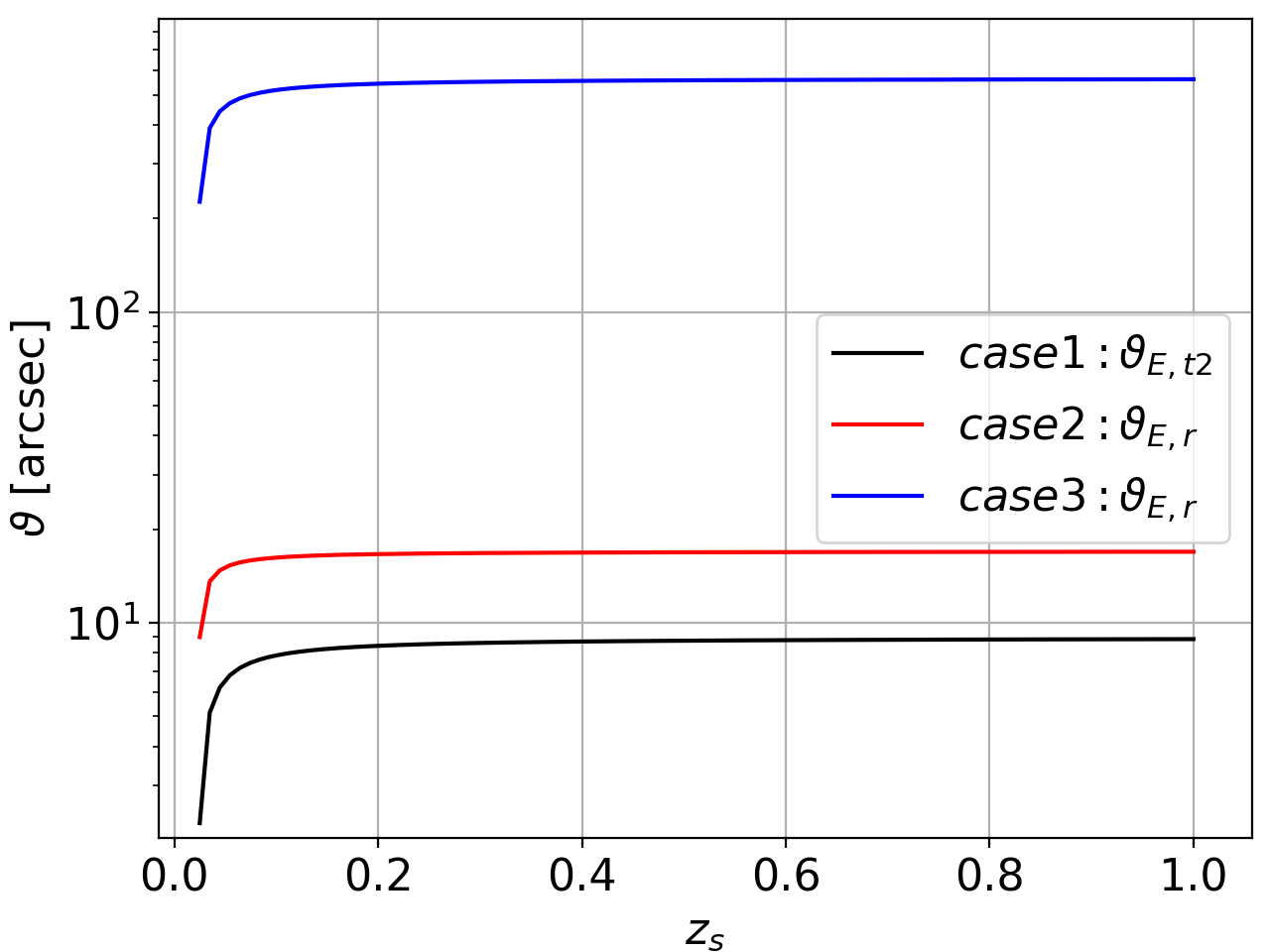} \hfill
\includegraphics[width=0.45\textwidth]{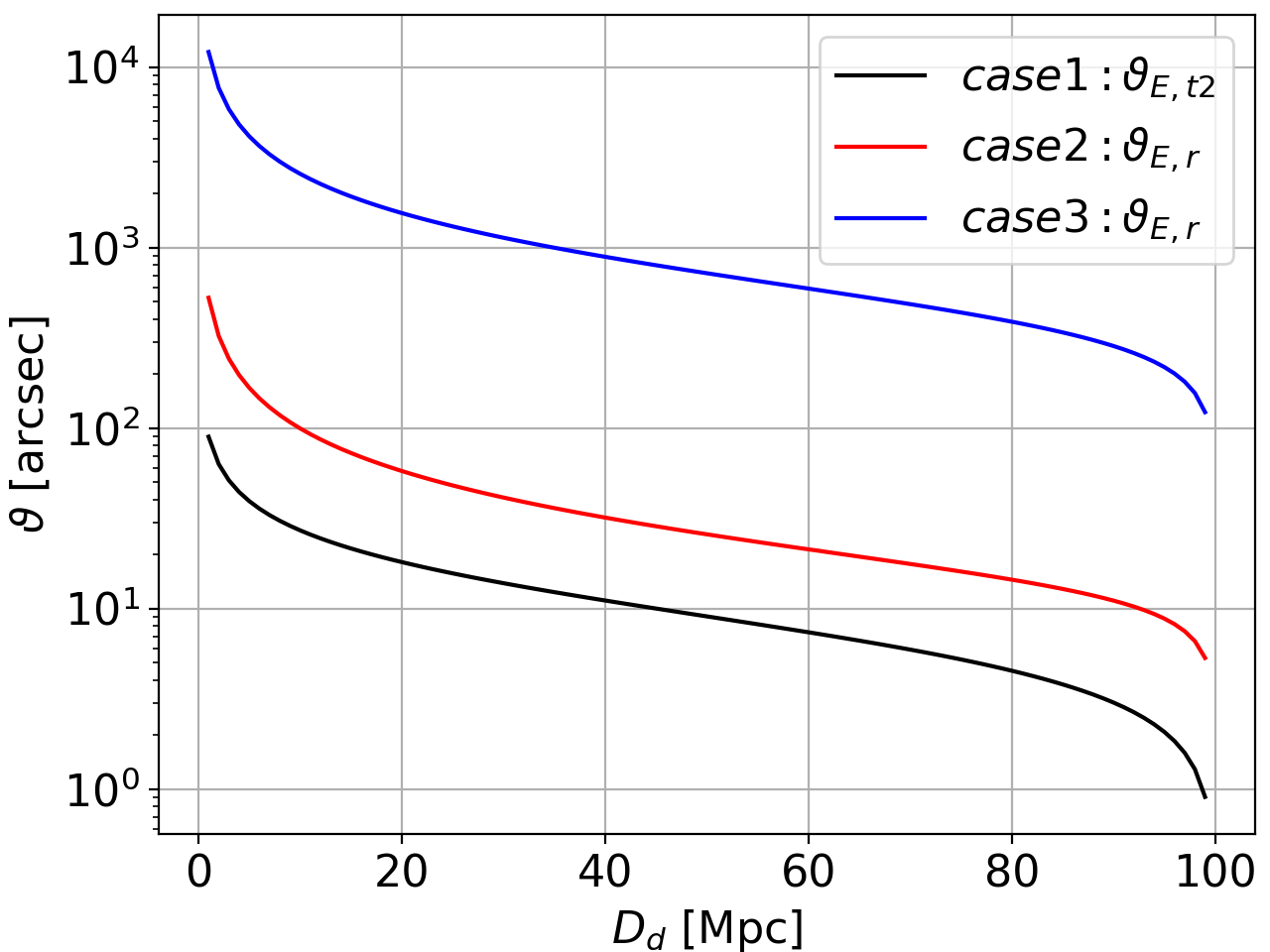}
\caption{Left: Lensing signals for scenario one. Right: Analogous plot for scenario two. Details see text, $\vartheta_\mathrm{E,t2}$ denotes the largest Einstein radius for case~1, $\theta_\mathrm{E,r}$ denote the extent of radial giant arcs for cases~2,3 due to the lack of an Einstein ring.}
\label{fig:lensing}
\end{figure}

To estimate the extent of the light deflections, we investigate two different scenarios: 
In the first, we place the object at $D_\mathrm{d} = 100$Mpc as the outmost possible position in the linear Hubble flow and assume a $\Lambda$CDM-like cosmology to investigate lensing effects on cosmic scales for sources at typical lensing redshifts $z_\mathrm{s}=0.05$ to 1.0 (this configuration is also assumed in Fig.~\ref{fig:rings}, right).
In the second scenario, we place the background source at $D_\mathrm{s} = 100$Mpc and investigate the lensing effects for an object at distances $D_\mathrm{d}=1$ to 99Mpc, independent of cosmological assumptions. 
Fig.~\ref{fig:lensing} (left) shows the size of the largest Einstein ring or the radial arcs for scenario one, Fig.~\ref{fig:lensing} (right) shows the results for scenario two. 

For comparison, galaxy-scale gravitational lenses have Einstein radii of about 1 arcsecond, Einstein rings of galaxy-groups and clusters are about one order of magnitude larger. 
The largest effective Einstein radius found so far is $55$~arcseconds and was observed for a galaxy cluster of mass $7.4\times 10^{14} \, M_\odot$ in \cite{bib:Zitrin}.
Since the size of the largest Einstein radius for case~1 is about 10 arcseconds, on average and additional dark and luminous matter may contaminate the observables as well, the clearest signature of a naked singularity is the sub-arcsecond additional Einstein ring which is lacking for any other (strong) gravitational lens, black hole, or even a weakly naked singularity without an event horizon but a still existing photon sphere. 
The objects with higher masses are even harder to identify, which implies that these objects might exist in our universe and have not been detected yet, as observable signatures are ambiguous and much less extreme than intuitively imagined.

Comparing the signatures of our charged super-extremal black holes with primordial stupendously large Schwarzschild black holes discussed by \cite{bib:Carr2021}, our example cases still lack a generating process, but avoid causing Einstein radii much larger than 500 arcseconds. 
A statistical forecast might reveal that the probability of such an observation is so low that existing constraints from non-detections are consistent with this kind of objects as well. 
However, the possible electro-magnetic effects in the vicinity of a Schwarzschild black hole are decoupled and there is no second Einstein ring for such objects, either. 
So, altogether, these ordinary black holes are even harder to identify.

\subsection*{Conclusion}

The Cosmological Constant, $\Lambda$, in Einstein's field equations can find a unique interpretation as the reference background energy-momentum density that separates energy-momentum content causing space-time curvature from a homogeneous and isotropic inert background. 
Depending on the prior constraints on the space-time metric and the energy-momentum tensor based on the matter and radiation content observed in our universe, it is possible to infer $\Lambda$ from Einstein's field equations and the data. 
As motivated in this essay, this interpretation can be valid on all scales, not only when employing Einstein's field equations to cosmology. 
Consequently, this interpretation can be tested in near-field cosmology, for instance, when constraining $\Lambda$ in binary systems in our local neighbourhood. 
Bounds from soon to be obtained data will be able to detect a non-zero $\Lambda$. 

In order to find the still unknown material source of the small but non-zero $\Lambda$ on cosmic scales, we can look for so-far undetected objects in the local, late-time universe, as well. 
Subsequently, these objects can be coarse-grained to a continuous background density that should at least partly explain the cosmic value for $\Lambda$.  
Very massive black holes, most likely of primordial origin seem to be promising candidates. 
We chose non-rotating super-extremal Reissner-Nordstr\"{o}m black holes with charge-to-mass-ratios that turn these objects into strongly naked singularities. 

Contrary to the first impression, these objects could cause disruptive, catastrophic effects, we showed that the far field of these extreme structures generates rather moderate perturbative phenomena and that their extreme nature actually avoids creating observable signatures in disagreement with current observations. 
Hence, even though it is still unclear how such naked singularities are generated, we cannot reject the hypothesis that they actually exist and are the material source of dark energy. 

The particular imprints in observables like rotation measures and strong gravitational lensing effects may be detected in sky surveys in radio and optical wave bands with current and upcoming telescopes. 
It is also possible that existing observations challenging the Cosmological Principle as summarised in \cite{bib:Aluri2022} already contain signatures of such objects and only the Cosmic Censorship conjecture has impeded us from investigating this hypothesis further.   
Facing the current tensions in our cosmological concordance model, not knowing much of the material constituents of the necessary dark sector, and only starting to understand the impact of electro-magnetic fields on structure creation and growth, further research on (primordial) black holes and potentially existing naked singularities looks like a promising path to resolve tensions with a minimum extension of currently known particles and physical laws. 
With more quantitative constraints from data, we may also improve on the currently rather arbitrary estimates about charge-to-mass ratios, mutual distances, and other prior assumptions that were required to obtain our first results. 
\vfill
\noindent
\textbf{Image credits: \\ Title page: Craiyon, all others: own work.}

\subsection*{References}
\begingroup
\renewcommand{\section}[2]{}%

\bibliographystyle{spmpsci}      
\bibliography{lambda}   
\endgroup

\end{document}